\documentclass{aa} 

\usepackage{graphicx}
\usepackage{txfonts}
\usepackage{orcidlink}
\usepackage{appendix}
\usepackage{natbib} 
\usepackage{comment}
\usepackage{subfigure}
\usepackage{subcaption}
\usepackage{supertabular}

\newcommand{\OIa}{[O\,{\scriptsize I}]\,$\lambda$6300~}
\newcommand{\OIb}{[O\,{\scriptsize I}]\,$\lambda$5577}

\newcommand{\SIIa}{[S\,{\scriptsize II}]\,$\lambda$6731~}

\begin{document}

   \title{Variability of the DG Tau Forbidden Emission Line Low Velocity Component   \thanks{Based on observations made with ESO Telescopes at the La Silla Paranal Observatory under programme IDs 384.C-0821(B) and 108.22M8.001 and on data collected by High Dispersion Spectrograph (HDS) at Subaru Telescope which is operated by the National Astronomical Observatory of Japan. We are honored and grateful for the opportunity to observe the Universe from Maunakea, which has cultural, historical, and natural significance in Hawaii. }}

\titlerunning{DG Tau LVC Variability}

\newcommand{\oiI}{[OI]$\lambda$6300 }
\newcommand{\oiII}{[OI]$\lambda$6363 }
\newcommand{\oiIII}{[OI]$\lambda$5577 }
\newcommand{\siiI}{[SII]$\lambda$6717}
\newcommand{\siiII}{[SII]$\lambda$6731 }
\newcommand{\siiIII}{[SII]$\lambda$4068 }
\newcommand{\niiI}{[NII]$\lambda$5755 }
\newcommand{\niiII}{[NII]$\lambda$6583 }
\newcommand{\LiI}{Li$\lambda$6707 }
\newcommand{\HeI}{HeI$\lambda$5876 }
\newcommand{\NaD}{NaD}

\newcommand{\Msun}{M$_{\odot}$}
\newcommand{\Lsun}{L$_{\odot}$}
\newcommand{\Tsun}{T$_{\odot}$}
\newcommand{\Mstar}{M$_{\star}$\ }
\newcommand{\Lstar}{L$_{\star}$\ }
\newcommand{\Rstar}{R$_{\star}$\ }
\newcommand{\Tstar}{T$_{\star}$\ }
\newcommand{\Rsun}{R$_{\odot}$}
\setlength{\parindent}{0pt} 

   \author{N. Otten
          \inst{1,2}
          \orcidlink{0000-0002-2530-4137} 
          E. Whelan
          \inst{1}
          \orcidlink{0000-0002-3741-9353}
          Y.-R. Chou
          \inst{3,4}
          \orcidlink{0009-0004-9608-6132}
          M. Takami
          \inst{3}
          \orcidlink{0000-0001-9248-7546}
          A. Murphy
          \inst{3}
          \orcidlink{0000-0003-0376-6127}
          and A. Banzatti
          \inst{5}
          \orcidlink{0000-0003-4335-0900}}

    \institute{Department of Physics, Maynooth University, Maynooth, Co.Kildare, Ireland \\
     \email{noah.otten@mu.ie}
         \and   
        European Southern Observatory,
        Alonso de Córdova 3107,
        Vitacura, Casilla 19001,
        Santiago de Chile,
        Chile
         \and
        Institute of Astronomy and Astrophysics, Academia Sinica, 11F of Astronomy-Mathematics Building, No.1, Sec. 4, Roosevelt Rd, Taipei
        10617, Taiwan, R.O.C
         \and
        Max Planck Institute for Astrophysics, Karl-Schwarzschild-Straße 1, 85748 Garching bei München, Germany
        \and
        Department of Physics, Texas State University, 749 N Comanche Street, San Marcos, TX 78666, USA}

   \date{Received -- ; Accepted --}

    \abstract
    {Optical Forbidden Emission Lines (FELs) come from transitions with long radiative decay times ($\approx$ 100s) that require low density environments where collisions between atoms are rare. They are produced in the low density gas found in the outflows (jets and winds) driven by low mass young stellar objects (YSOs) and frequently reveal distinct velocity components within the outflow including a so-called Low Velocity Component (LVC). A question pertinent to the removal of excess angular momentum in YSOs is whether the LVC traces a magneto-hydrodynamic wind (MHD) or a photevaporative wind. Here the jet and LVC of the classical T Tauri Star DG~Tau is studied. The velocity of the DG~Tau jet has been decreasing since 2006 making it a particularly interesting source for this work.}
    {The aim is to investigate the connection between the high velocity jet and the LVC in DG~Tau and to better understand the origin of the LVC by examining spectral and spatial changes over time. }
    {Kinematic fitting and spectro-astrometry are applied to  three epochs of high spectral resolution data spanning $\approx$ 18 years to conduct a detailed study of the changes in the LVC over time.}
    {A decrease in velocity of $\approx$ 100 kms$^{-1}$ from 2003 to 2021 is in agreement with the known slowing of the DG~Tau jet. The kinematic fitting of the [O I]$\lambda$6300, \OIb~and \SIIa lines over the three epochs of data reveal the complex nature of the optical FELs. In agreement with \cite{Chou2024} up to six blue-shifted components in the FEL line profiles alongside a red-shifted wing are identified. The three observed LVC sub-components (LVC-\textbf{H}igh, LVC-\textbf{M}edium and LVC-\textbf{L}ow) are consistent with entrained jet material, a disk wind and a dense upper disk atmosphere respectively. Despite the strong variability of the jet components over the three epochs, the LVC is found to be far more stable and only the relative brightness of the three LVC sub-components is seen to change. A constraint of $\geq$ 2 au is put on the minimum de-projected height of the LVC-M in \OIb\ where there is no contribution from the jet. }
   {The results support a disk wind origin for the LVC-M sub-component but cannot distinguish between a photoevaporative or MHD wind origin. The minimum \OIb\ LVC-M height of $\geq$ 2 au indicates that this wind is launched inside the gravitational potential well of DG~Tau and favors a MHD wind origin for the LVC-M. That the peak velocity of the LVC-M does not change significantly needs to be investigated further in the context of a common origin for jets and MHD disk winds. Future studies would benefit from higher spectral resolution data to reduce blending between the outflow components and higher cadence sampling in time to explore a time lag between changes in the jet and the LVC.}


   \maketitle
%
\section{Introduction} 
Classical T Tauri stars (CTTSs) are class II low mass YSOs and the precursors of protoplanetary systems. They are active accretors and drive jets and winds \citep{Ray2007}. As a result, their spectra show an abundance of emission lines tracing accretion and outflow activity \citep{Whelan2014a}. From early in the study of CTTSs their FEL profiles were known to be multi-component \citep{Hartigan1995} with a frequently spatially extended high velocity component (HVC, |v| $> 100~kms^{-1}$) tracing the collimated jet  \citep{Hirth1997} and a spatially compact low velocity component (LVC, |v| $< 40~kms^{-1}$) tracing a wind \citep{Kwan1988}. This wind could be a magneto-hydrodynamic (MHD) wind and therefore connected to the jet \citep{Pascucci2024, Birney2024} or a photoevaporative wind and therefore thermally driven \citep{Rab2023}. Recently, there has been renewed interest in the FEL LVC as a means of investigating if MHD winds solve the angular momentum transport problem in YSOs \citep{Pascucci2024}. Recent simulations \citep{Bai2016} that include non-ideal MHD effects have shown that low velocity winds in the inner disk are critical in providing the spin-down torques needed to remove excess angular momentum and hence, enabling stellar mass accretion. Finding an observational tracer of these winds and testing the findings of these simulations is critical in understanding angular momentum transport in YSOs. The FEL LVC is a promising potential tracer of these low velocity winds and as such, may be important in the removal of excess angular momentum in these young stellar systems. However, the compact nature of the LVC presents an observational challenge to understanding its origin. Two methods have been used to address the compact nature of the LVC and study it. They are kinematic fitting of the line profiles through Gaussian decomposition \citep{Banzatti2019} and spectro-astrometry which recovers spatial information from compact emission line regions \citep{Whelan2008}. \\

\begin{figure*}[tbh]
    \centering 
\begin{minipage}[t]{\textwidth}
\begin{subfigure}{}
  \includegraphics[width=5.8cm,height= 6.7cm, trim= 0cm 0cm 0cm 0cm, clip=true]{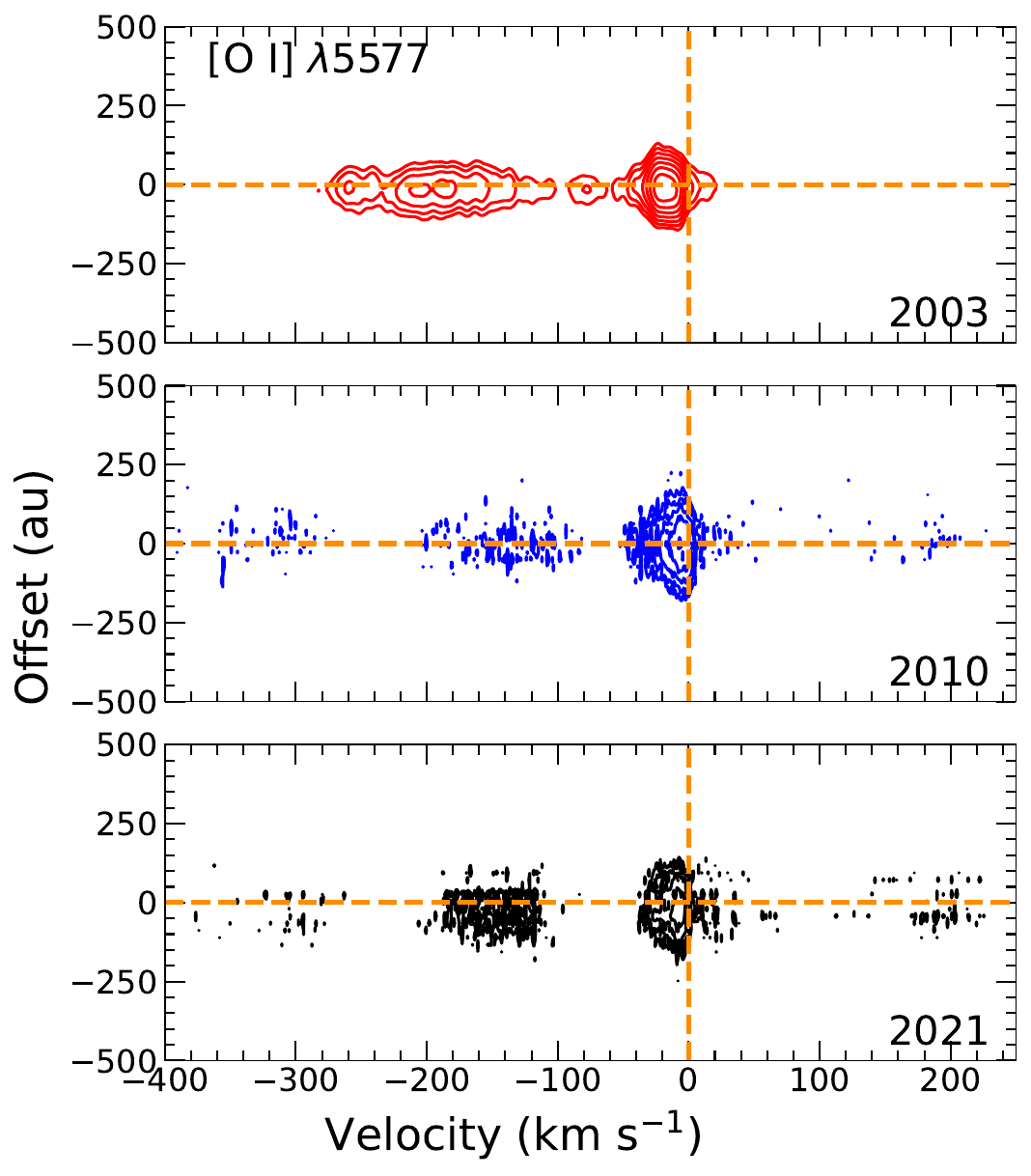}
\end{subfigure}
\begin{subfigure}{}
  \includegraphics[width=5.7cm,trim= 0.5cm 0cm 0cm 0cm, clip=true]{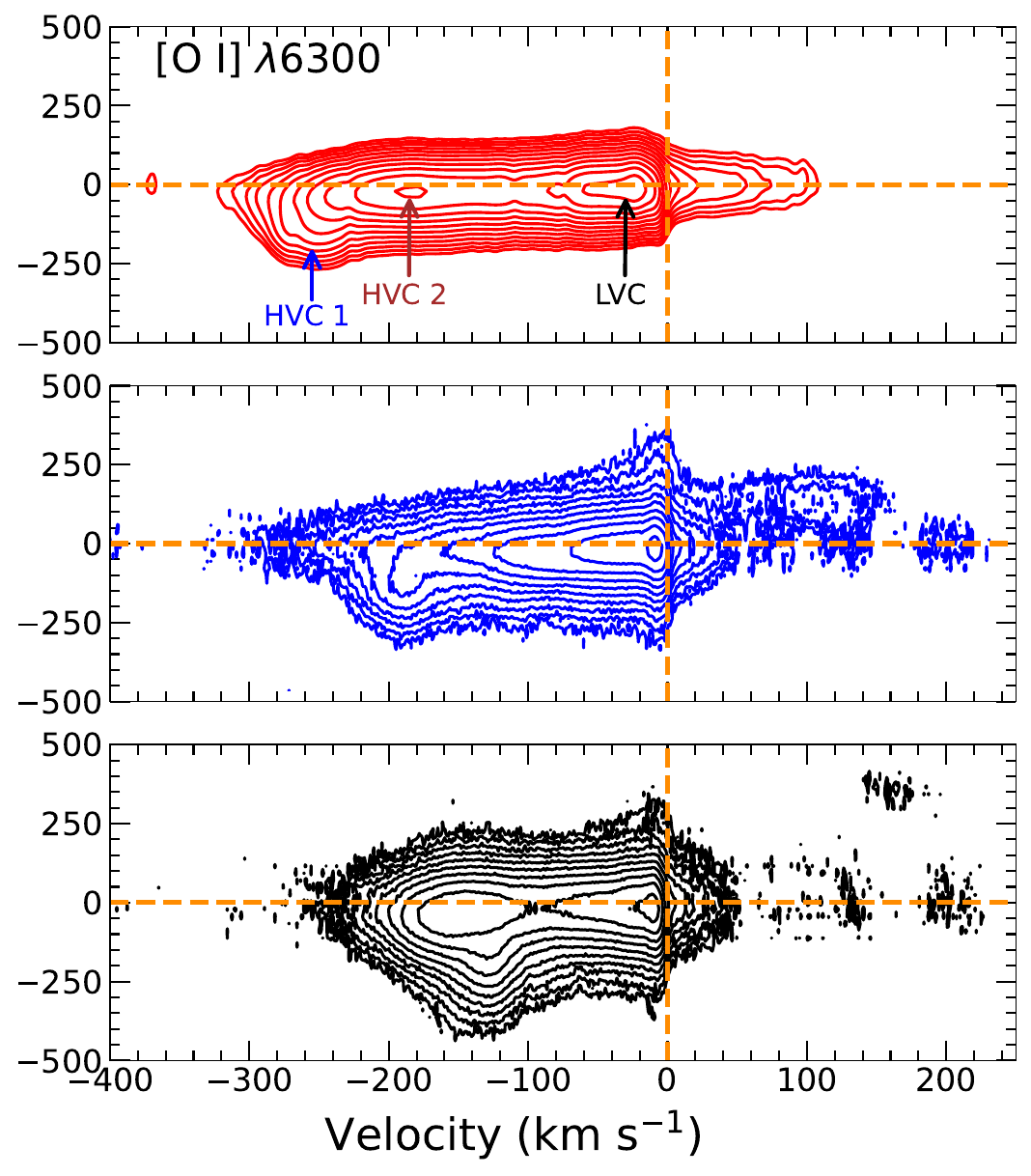}
\end{subfigure}
\begin{subfigure}{}
  \includegraphics[width=5.7cm,trim= 0.5cm 0cm 0cm 0cm, clip=true]{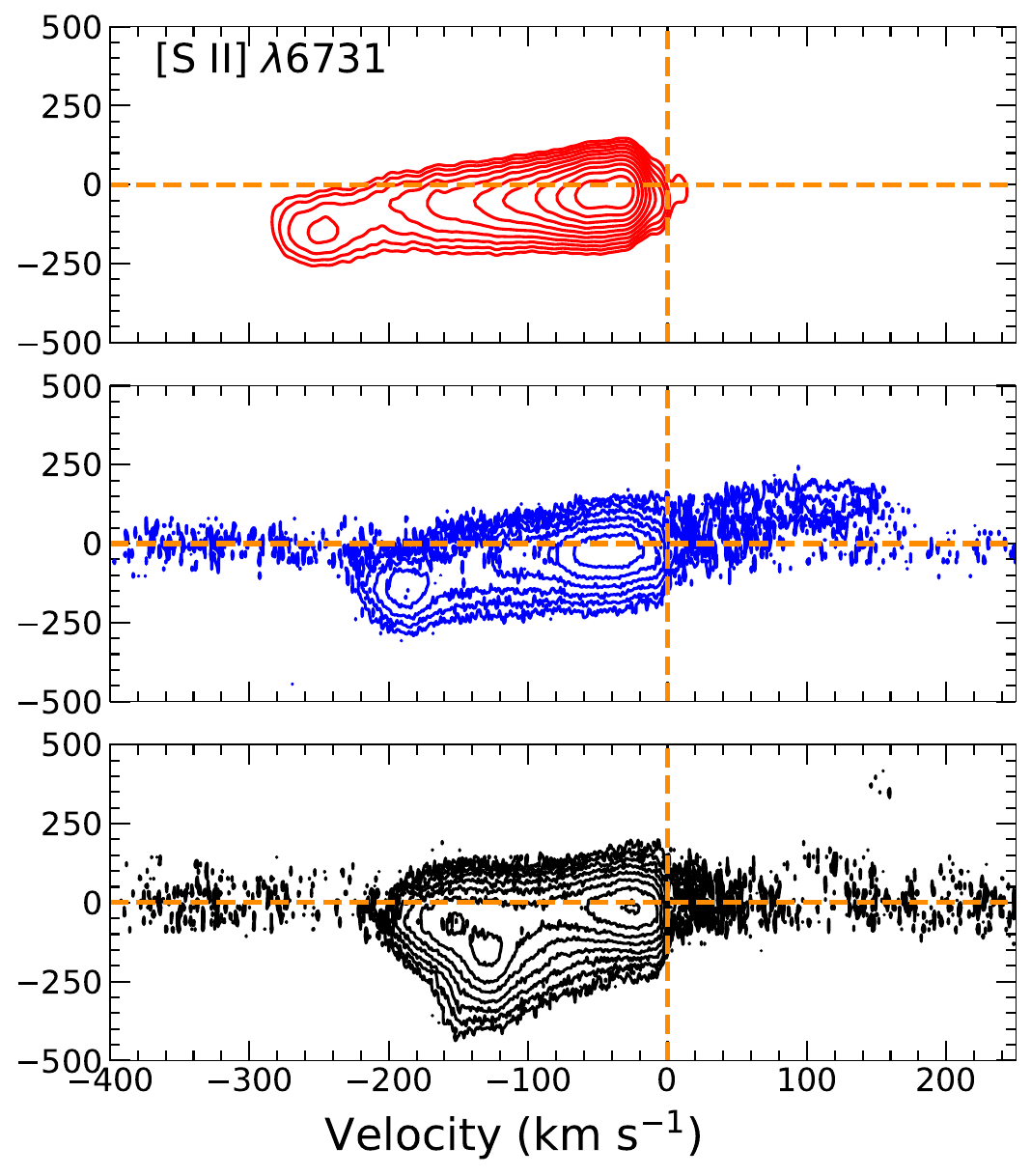}
\end{subfigure}
\end{minipage}
\caption{Position velocity diagrams of the  \OIb~(left) \OIa (middle) and \SIIa (right) FELs in 2003, 2010 and 2021 from DG~Tau. Contours start at $3\sigma$ (2003) and $7\sigma$ (2010, 2021) above the background rms noise and increase by factors of $\sqrt{2}$ (2003) and $\sqrt{3}$ (2010, 2021). The kinematic components identified by \cite{Chou2024} in the 2003 data (HVC1, HVC2, LVC) are marked. The assumed distance to DG~Tau is 125~pc (see Table \ref{properties_table}).}
\label{DGTau_pvplots}
\end{figure*}

Previous literature regarding the LVC (e.g. \cite{Simon2016,McGinnis2018,Fang2018,Nisini2024}) have used kinematic fitting to show that the LVC can be further divided into sub-components. These studies observed both a broad and narrow LVC sub-component in FEL profiles. These broad and narrow sub-components are separated on the basis of their measured Full Width at Half Maximum (FWHM). The broad sub-component, referred to as the LVC-BC has a FWHM $\geq$ 40 kms$^{-1}$ whereas the narrow sub-component, referred to as the LVC-NC has a FWHM $\leq$ 40 kms$^{-1}$. 
\cite{Fang2018} found statistical correlation between the \OIa line luminosity of the HVC and LVC to the stellar accretion luminosity in a sample of 48 stars. They found that the LVC-BC is more tightly correlated to the stellar accretion luminosity than the LVC-NC. They also find the ratio of \OIb~to~\OIa differs between the LVC-BC and the HVC. These findings combined corroborate the idea that the LVC-BC traces a hot/dense region of gas close to the base on an MHD disk wind. 
\newline\par\noindent
\cite{Banzatti2019} expanded the work of \cite{Simon2016} and \cite{Fang2018} to a sample of 65 stars. They find correlations between the kinematics of the LVC (centroid velocity and FWHM) to the equivalent width of the HVC and the stellar accretion luminosity. This work reinforces the idea that the LVC-BC traces the dense/hot base of an MHD disk wind. The origin of the LVC-NC remains a matter of debate. Previous literature indicates that the LVC-NC may be part of an MHD wind \citep{Fang2018,Banzatti2019} or that the LVC-NC sub-component can be produced by photoevaporative winds \citep{Rab2023}. A key deficiency of these kinematic fitting studies of many young stars is the lack of spatial information. The poor constraint on the vertical extent of the emission traced by these LVC sub-components above the disk mid-plane limits the accuracy of the derived mass outflow rates. The LVC is compact so measuring the height of the LVC emission above disk mid-plane presents a challenge. The technique of spectro-astrometry can be used to address this and provide critical constraints on the vertical extent of the emission traced by these LVC sub-components. 
\newline\par\noindent

\begin{figure*}
\centering
\includegraphics[width=19cm]{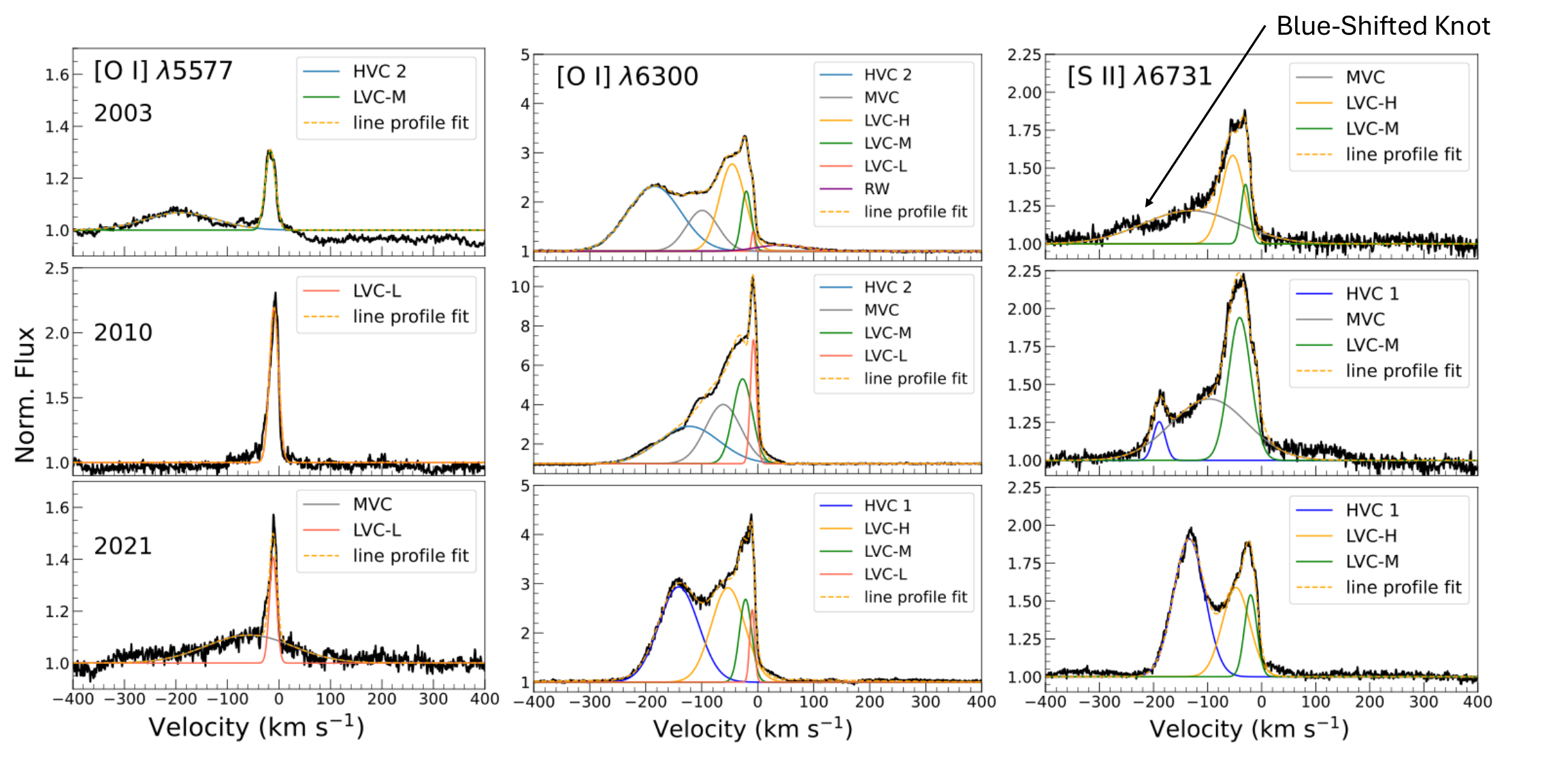}
    \caption{Kinematic fitting comparison of the \OIb, \OIa and \SIIa line profiles via Gaussian decomposition in 2003, 2010 and 2021. The line profiles are shown in the stellocentric frame and have been normalised to the continuum level. These profiles are extracted over the full spatial range of the PV diagrams shown in Fig.\ref{DGTau_pvplots}.} 
        \label{kinematic_fitting}
\end{figure*}

For example, \cite{Whelan2021} applied spectro-astrometry to optical spectra from the star RU~Lupi in order to study its FEL LVC. Spectro-astrometry involves the fitting of a Gaussian to spatial profiles extracted perpendicular to the dispersion axis of a 2D spectrum. The peak position of each fit to a spatial profile is called a spatial centroid. These spatial centroids can be plot as a function of velocity and the result is referred to as a position spectrum. Position spectra measure the centroid offset of the brightest extended emission traced by an emission line region relative to the position of the star. The work of \cite{Whelan2021} provided better constraints on the the vertical extent of emission above the disk mid-plane compared to the previous kinematic fitting studies. Their results show a negative velocity gradient in region of the position spectrum related to the LVC-NC sub-component. A negative velocity gradient in a position spectrum refers to an increase in the recorded position centroid offset with decreasing velocity. Physically it means a decrease in the velocity of the emission traced by the spectro-astrometry in the forbidden emission line region as height above the disk mid-plane increases. Modelling has shown that this negative gradient is indicative of a disk wind \citep{Ponto2011}. The spectro-astrometry of \cite{Whelan2021} traced the extent of the LVC-NC to 2 au. This is within the gravitational potential well of RU~Lupi. This is evidence that this LVC-NC sub-component may be tracing an MHD disk wind. 
\newline\par\noindent
Conversely, a positive velocity gradient in the position spectrum refers to an increase in the recorded position centroid offset with increasing velocity. Physically it means an increase in the velocity of the emission traced by the spectro- astrometry in the forbidden emission line region as height above the disk mid-plane increases. This positive gradient is as feature of jets \citep{Hirth1997} and close to the source it is indicative of the jet launching process. At greater distances from the source this positive velocity gradient could be caused by entrainment or material ejected at higher velocities accelerating slower jet material \citep{Davis2001,Whelan2004}. Eventually, for the high velocity region, the spectro-astrometric signal changes so that a decrease in offset is recorded with velocity and this often coincides with knots in the jet (see Figures 3 and 5 of \cite{Whelan2004}). These positive velocity gradients have also been observed in previous works that have used spectro-astrometry \citep{Whelan2024}. 
\newline\par\noindent
Many CTTS jets show kinematic variability which could be connected to variable accretion \citep{Pyo2024, Takami2023}. If the LVC traces a MHD wind then it is reasonable to expect that it will vary with a variable HVC. The disk wind model for jet launching purports that the jet is the inner part of the larger MHD disk wind which becomes collimated by magnetic fields \citep{Ferrreira2006}. Therefore, additional information on the origin of the LVC could be gleamed by studying variability. Variability of the LVC has been noted previously in the literature \citep{Simon2016,Nisini2024}however, no dedicated study of the variability of the LVC has been conducted to date. This article is the first dedicated study of variation in the LVC over time. 
\newline\par\noindent
We note here that \cite{Takami2023} report a factor of two decrease in the tangential velocity of the DG Tau blue-shifted jet over a twenty year period pointing to a remarkable change in the velocity at which DG~Tau ejects material. This makes DG~Tau a particularly interesting object in which to study the LVC. The goal of the work presented here is to determine if spectral and spatial changes in the DG Tau HVC are accompanied by corresponding changes in the LVC. 
\newline\par\noindent
This article expands on the previous literature and presents a dedicated study of the variability in the blue-shifted LVC sub-components of DG~Tau for three optical FELs. We use kinematic fitting and spectro-astrometry, applied to three epochs of high resolution spectra covering the FELs \OIb, \OIa and [S\,{\scriptsize II}]\,$\lambda$6731. These lines are selected because they have very different critical densities, decreasing from $\sim$ 10$^{8}$ cm$^{-3}$ for \OIb\ to $\sim$ 10$^{4}$ cm$^{-3}$ for [S II]$\lambda$6731, hence they trace different parts of the outflow system \citep{Giannini2019}.   
\newline\par\noindent
Recent work by \cite{Chou2024} use kinematic fitting and spectro-astrometry to isolate six different blue-shifted emission components in the FELs of DG Tau. They found two HVCs (HVC1, HVC2), a medium velocity component (MVC) and three low velocity components, (LVC-\textbf{H}igh, LVC-\textbf{M}edium and LVC-\textbf{L}ow) for high, medium and low velocities within the LVC. Hereafter, we refer to these LVC sub-components as the LVC-H, LVC-M and LVC-L respectively. They argue that the HVC1 and HVC2 trace the extended jet and a stationary shock at the jet base respectively. \cite{Chou2024} also use line ratios and modelling to conclude that the LVC-H, LVC-M and LVC-L are associated with an interaction region between the jet and disk wind, a disk wind and the upper disk atmosphere respectively. Their data is the first epoch of data used for our time variability study. As such, for consistency we refer to the velocity components observed in this article using the nomenclature of \cite{Chou2024}. The rest of the article is organised as follows: Section \ref{Sec2} outlines the observations undertaken and the data reduction and analysis steps. In section \ref{Sec3} we present the PV diagrams and the results of our kinematic and spectro-astrometric analysis. Lastly, in section \ref{Sec4} we discuss our findings and draw our conclusions from them.

\section{Observations, data reduction and analysis}\label{Sec2}

\begin{table}
\caption{Properties of DG~Tau. }             
\label{properties_table}      
\centering                          
\begin{tabular}{c c c}        
\hline\hline                 
Property & Value & Reference \\    
\hline                        
   SpTy& K7 & B19  \\      
   Mass (\Msun) & 0.80 & B19 \\
   Radius (\Rsun) & 1.85 & G16  \\ 
   Age (yrs.) & 3$\times10^{5}$ & B90 \\
   log($\dot{M}_{\rm acc}$) (M$_\odot$/yr) & -8.39 &  HH08 \\
   Disk Type & full & P15 \\
   Disk PA ($^\circ$) & 119 $\pm$ 24 & I10   \\
   Blue Jet PA ($^\circ$) & 226 $\pm$ 1 & L16 \\
   Disk incl. ($^\circ$) & 32 $\pm$ 2 &  B19 \\ 
   $A_{v}$ (mag) & 1.6 $\pm$ 0.15 & HH14 \\
   Distance (pc) & 125.3 & GDR3  \\
   $V_{RV}$ ($kms^{-1}$) & 19.0 $\pm$ 0.5 & this work  \\

\hline                                   
\end{tabular}
\tablefoot{B19 = \cite{Banzatti2019}; G16 = \cite{Grankin2016}; B90 = \cite{Beckwith1990}; HH08 = \cite{Herczeg2008}; P15 = \cite{Pascucci2015}; I10 = \cite{Isella2010}; L16 = \cite{Liu2016}; HH14 = \cite{Herczeg2014};  GDR3 = GAIA Data Release 3 \citep{Vallenari2023,Prusti2016}. }  
    \label{dgtauprops}
\end{table}

\begin{table}[hb]
\caption{Observation Log}             
\label{obslog}      
\centering                          
\begin{tabular}{c c c}        
\hline\hline                 
Date & Slit PA & Exposure Times \\    
(dd-mm-yyyy) & ($^\circ$) & (ncycles $\times$ s) \\
\hline                        
   06-12-2003 & 44.4 & 4 $\times$ 900 \\
   06-12-2003 & 224.4 & 4 $\times$ 900 \\
   07-12-2003 & 134.4 & 4 $\times$ 600 \\
   07-12-2003 & 314.4 & 4 $\times$ 600 \\
   07-01-2010 & 46 & 2 $\times$ 755 $^{*}$\\
   14-11-2021 & 46 & 1 $\times$ 300 \\
   14-11-2021 & 136 & 1 $\times$ 300 \\
   16-11-2021 & 46 & 3 $\times$ 800\\
   13-02-2022 & 136 & 3 $\times$ 800 \\
   
\hline                                   
\end{tabular}
\tablefoot{$^{*}$ = Only one exposure was used as the source had moved in the slit during the second exposure.}
\end{table}

DG~Tau (Table \ref{dgtauprops}) was observed in 2003 with SUBARU/HDS \citep{Chou2024} and in 2010/2021 with the Ultraviolet and Visual Echelle Spectrograph (UVES) \citep{Dekker2000} on the Very Large Telescope (Table \ref{obslog}). The instrument slits were parallel to the known jet position angle. The details regarding the reduction of the 2003 data is presented in \cite{Chou2024}. The UVES data were reduced using the UVES pipeline (version 6.1.6) \citep{esoreflex2013} to produce 2D spectra. PV diagrams were constructed (Figure \ref{DGTau_pvplots}) from the full spatial range of these 2D spectra after continuum subtraction. 
\newline\par\noindent
The continuum subtraction was preformed on each 2D spectrum of the \OIb,~\OIa and \SIIa lines following this procedure. A \textbf{\textit{Python}} script begins by masking the regions that contain emission or absorption features along each dispersion axis in the spectrum. Following this, a polynomial is fit to the unmasked regions to obtain an estimate of the continuum along each dispersion axis. The estimated continuum is then subtracted from the 2D spectrum in a line-by-line fashion to remove the stellar contribution from the spectrum.  
\newline\par\noindent
1D spectral line profiles were then extracted from these 2D continuum subtracted spectra. The region in which the line profiles are extracted covers the whole spatial range of the 2D data. The line profiles were then corrected for telluric and photospheric absorption features using the ESO MOLECFIT pipeline \citep{Smette2015,Kausch2015} and the photospheric correction method outlined in \cite{Hartigan1989}. The kinematic fitting of the 1D spectra was then carried out following the procedure of \cite{Chou2024}. This method involves fitting a composite model of independent Gaussian components to reproduce the line profile. The number of required Gaussian components for the model is determined via a reduced chi-square minimisation. Additional Gaussian components are added to the model only if the reduced chi-square improves by at least a factor of 20\% compared to the composite model with one fewer Gaussian components. 
\newline\par\noindent
All velocities quoted in this work are in the stellocentric frame. In order to convert the line profiles from the topocentric frame into the stellocentric frame two velocity corrections were applied to the line profiles. The stellar radial velocity correction was calculated following \cite{Pascucci2015}. The heliocentric velocity correction was calculated using the \textit{\textbf{radial\_velocity\_correction}} function available in \textit{\textbf{Astropy}} \citep{astropy:2013,astropy:2018,astropy:2022}. The \textit{\textbf{rvcorrect}} method available in the Image Reduction and Analysis Facility (IRAF) was also used to confirm the accuracy of the heliocentric velocity correction calculated using \textit{\textbf{Astropy}}. The uncertainties on the centroid velocity (V$_{c}$) and FWHM values obtained from the kinematic fitting were calculated following the method of \cite{Nisini2024}. These uncertainties were significantly less than the uncertainty calculated for the radial velocity of the source (0.5~kms$^{-1}$) so, this latter value is taken as the uncertainty on the kinematic fitting values.
\newline\par\noindent
Spectro-astrometry was applied to the 2D spectra before and after continuum subtraction as outlined in \cite{Whelan2021}. This was done to prevent the stellar brightness contribution contaminating the offset measured in the position spectrum for the brightest extended emission. The centroids measured for the extended emission before continuum subtraction are pulled back to the stellar position due to the stellar contribution to the 2D spectrum. The stellar continuum is subtracted as outlined in the previous paragraphs and the spatial centroids were remeasured less the continuum. The 1-$\sigma$ uncertainty on the recovered position of the extended emission depends on the signal to noise ratio (SNR) of the spatial profiles (see Equation \ref{sa_uncert} and \cite{Whelan2008}). In equation \ref{sa_uncert}, N$_{p}$ is the number of detected photons and the FWHM of the spatial profile is an approximate measure of the seeing. For the HDS observations anti-parallel slit position angles (PAs) were used to correct for spatial artefacts. For the UVES data anti-parallel slit PAs were not available and spatial  artefacts were ruled out by examining the spectro-astrometric position spectra in lines where no centroid offset is expected \citep{Whelan2009}.

\begin{equation}\label{sa_uncert}
    \sigma_{centroid} = \frac{seeing~(in~mas)}{2.3548 \times \sqrt{N_{p}}} \approx \frac{0.4 \times FWHM }{SNR}
\end{equation}

\begin{table*}[ht]
\caption{Kinematic fitting of the the blue-shifted outflow components in kms$^{-1}$. The centroid velocity and FWHM (in brackets) of each Gaussian are given for the HVC1, HVC2, MVC, LVC-H, LVC-M and LVC-L. The uncertainty on the centroid velocities and FWHM values 
is taken as  0.5~kms$^{-1}$. The spectra here are taken from the full spatial range of the PV plots shown in Fig. \ref{DGTau_pvplots}. The extended jet component (i.e. the HVC1, marked in \textbf{bold}) for the \OIa line in 2003/2010 and the \SIIa line in 2003 are identified as part of the blue-shifted wing and not identified as a separate component by the kinematic fitting. To measure these values a further spectrum was extracted from the extended jet region in 2003 and 2010.}
    \centering\tiny
    \begin{tabular}{l l l l l l l }
    \hline\hline 
  &HVC1  &HVC2  &MVC &LVC-H &LVC-M &LVC-L  \\ 
    \hline
    \textbf{\OIb} & & & & & &  \\
     2003 & - &  -191 (165) & - & - & -17 (23) & - \\
     2010 & - & - & - & - & - & -10 (21)  \\
     2021 & - &  & -58 (198) & - & - & -12 (17)\\
     \hline
     \textbf{\OIa} & & & & & & \\
     2003 & \textbf{-250 (50)} & -185 (112) & -99 (67) & -46 (54) & -20 (21) & -9 (6)  \\
     2010 & \textbf{-178 (82)} & -122 (126) & -62 (75) & - & -27 (41) & -8 (13) \\
     2021 & -141 (83) & - & - &  -53 (71) & -22 (26) & -10 (10)\\
     \hline
     \textbf{\SIIa} & & & & & &   \\ 
     2003  & \textbf{-249 (46)} & - & -131 (207) & -53 (49) & -30 (19) & -  \\
     2010  & -190 (26) & - & -97 (165) & - & -40 (49) & - \\
     2021 & -134 (68) & - & - & -48 (60) & -20 (27) & -  \\
     \hline
    \end{tabular} 
    \label{table_kinfit_comparison}
\end{table*}

\begin{figure*}
\centering\includegraphics[width=18cm,trim= 0cm 0cm 0cm 0cm, clip=true]{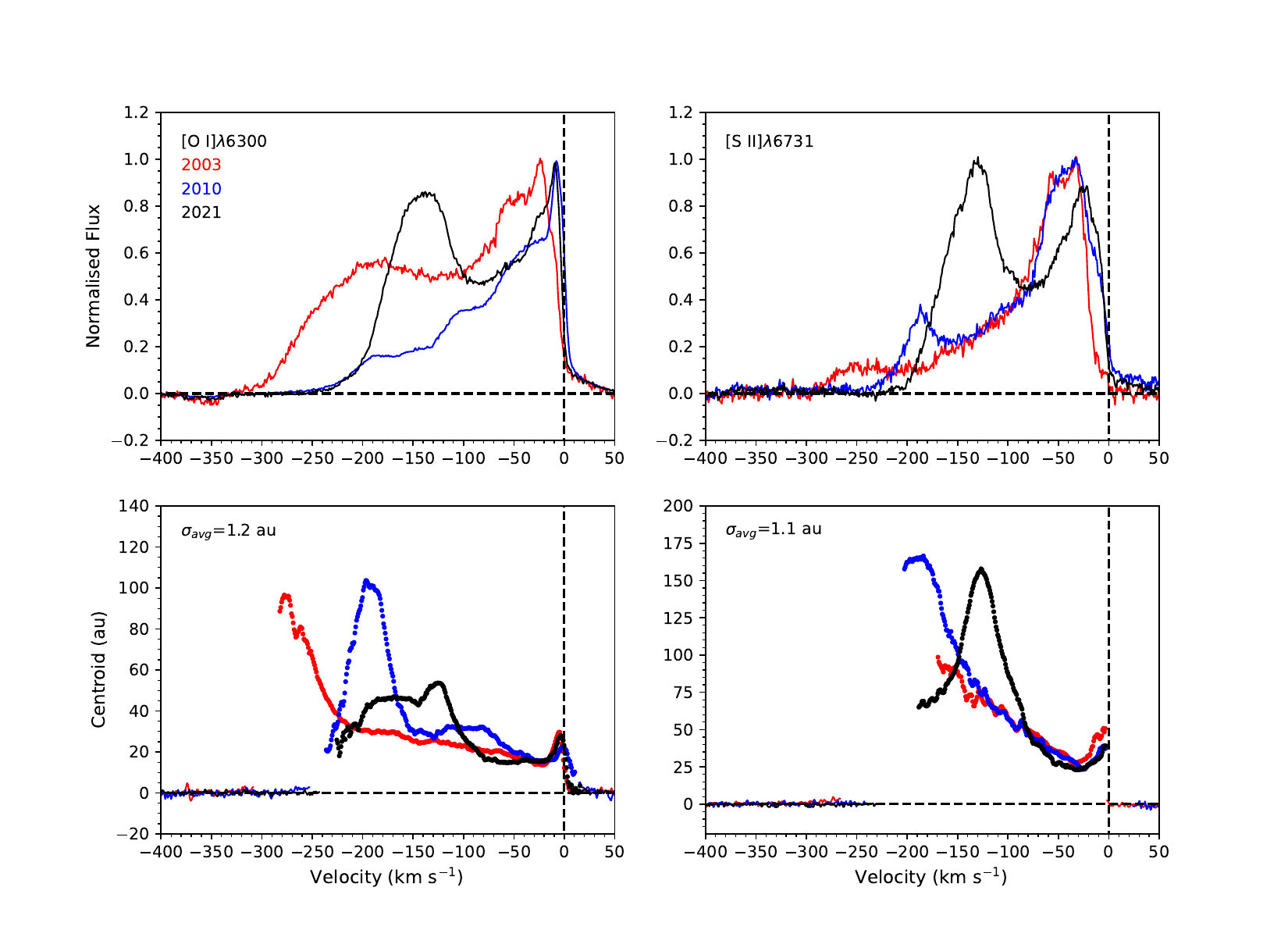}
    \caption{Top row: Comparison of the line profiles of \OIa and \SIIa for each epoch of data. The line profiles have been normalised to max peak height to emphasise changes in the line shape over time.  Bottom row: Position spectra for the \OIa and \SIIa lines in 2003 (red), 2010 (blue) and 2021 (black). The analysis of the \OIb~line is not presented here as it does not show an extended jet component due to its high critical density. } 
        \label{DGTau_SA_HVC}
\end{figure*}

\begin{figure*}
\centering\includegraphics[width=18cm,trim= 0cm 2cm 0cm 4cm, clip=true]{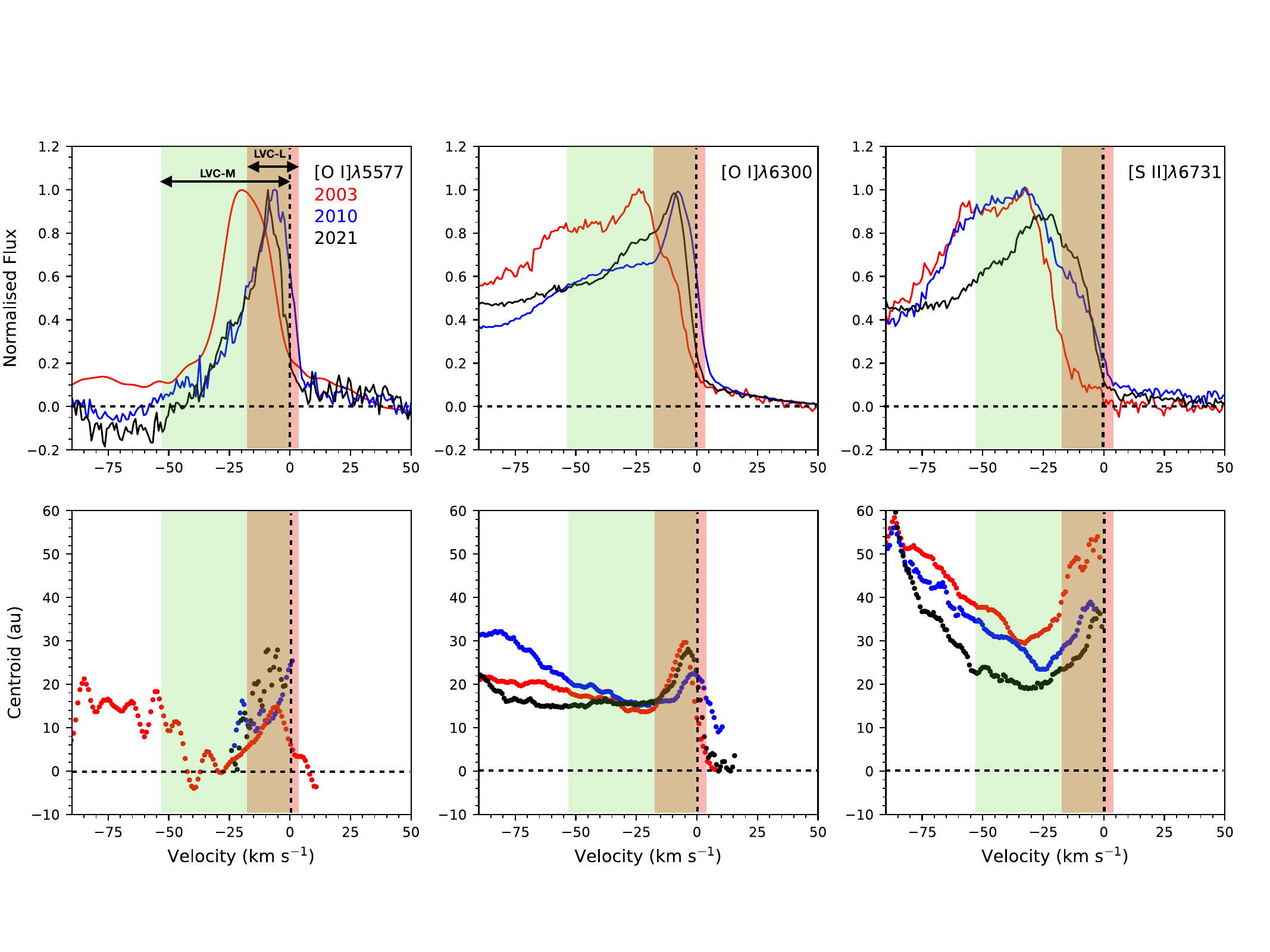}
    \caption{Top row: Comparison of the line profiles in the LVC region of each FEL in each epoch of data. 2003 = red, 2010 = blue and 2021 = black. These panels show a zoom of the line profiles shown in Fig. \ref{DGTau_SA_HVC} from -90 to 50 kms$^{-1}$. This is done to emphasise the change of line shape in the LVC region over time. The line profiles have been normalised to max peak height. Bottom row: The position spectra of the LVC region for each FEL in each epoch of observation. 2003 = red, 2010 = blue and 2021 = black. The 1-$\sigma$ uncertainties range from 1 to 5~au where 5~au is measured for the lowest signal to noise emission i.e. \OIb\ in 2021.} 
        \label{DGTau_1dSA_compare1}
\end{figure*} 


\section{Results}\label{Sec3}
Due to the relative faintness of the red-shifted outflow only the blue-shifted outflow was studied. Using the PV diagrams and kinematic fitting discussed in Section 2 (Table \ref{table_kinfit_comparison}, Figures \ref{DGTau_pvplots}, \ref{kinematic_fitting}) all the kinematic components first identified by \cite{Chou2024} are found but not in all lines or all years. Once identified they are investigated further using spectro-astrometry. 
 The differences between the emission line regions can be explained in terms of their different critical densities and the decreasing electron density of the outflow with distance from the star. \OIb\ traces the densest and therefore generally the most compact components. \OIa is the most complex with contributions from all the identified kinematic components. \SIIa traces the most extended emission. In Figures \ref{DGTau_SA_HVC} and \ref{DGTau_1dSA_compare1} for example, the \SIIa emission is extended up to twice the distance of the \OIa emission.  The HVC2 and LVC-L are not identified in the \SIIa line and \cite{Chou2024} argue that this is because the critical density of \SIIa is too low. 
The effectiveness of the kinematic fitting for identifying the different velocity components is limited by the relative brightness of the components and it should be noted that a non-detection by the kinematic fitting does not mean that a particular component is not present in the outflow. For example, even though the LVC-L is not detected in the kinematic fitting of the \SIIa line it does not mean that it is not present. Its contribution may be quenched due to critical density effects but it may still contribute to the overall line profile as described further in Section 3.2. The results for the 
the kinematic components associated with the jet by \cite{Chou2024} (HVC1, HVC2, MVC and LVC-H) and the LVC-L and LVC-M sub-components are now presented separately.

\subsection{Variability in the jet components}\label{Sec3.1}

As the HVC1 is part of the extended jet it can be studied through the PV diagrams and the kinematic fitting. It is always present in \OIa and \SIIa and its decrease in velocity by $\sim$ 100~kms$^{-1}$ from 2003 to 2021 follows the known decrease in the jet velocity over this period \citep{Takami2023,Pyo2024}. 
In the 2003 and 2010, in the extracted spectra shown in Figure \ref{kinematic_fitting}, it appears as a blue-wing and is not picked up by the kinematic fitting. To measure the 2003 and 2010 HVC1 velocity and FWHM for Table \ref{table_kinfit_comparison}, a second spectrum extracted from the jet region was used. The extended jet does not have a high enough excitation to produce detectable emission in [O I]$\lambda$5577 with the exposure times and equipment used, hence this component is not detected in the kinematic fitting or PV plots of this emission line. 
\newline\par\noindent
The HVC2 is detected in \OIa in 2003 and 2010, in both the PV plots and the kinematic fitting, but not in 2021. In 2021 it is likely blended with the HVC1 and MVC. The \SIIa line does not trace this component and \OIb\ traces the HVC2 in 2003 only. The decrease in the HVC2 velocity from 2003 to 2010, is in line with the decrease in the HVC1 velocity. 
\newline\par\noindent
The MVC and LVC-H are not always detected and only together in \OIa and \SIIa in 2003. This is probably due to blending between the components as the velocity of the jet decreases. \cite{Chou2024} associate the LVC-H with a jet/wind interaction region. They do not discuss the origin of the MVC but the decrease in velocity of this component from 2003 to 2021 may suggest that it is tracing newly ejected lower velocity knots which eventually become part of the extended jet. 
\newline\par\noindent
Figure \ref{DGTau_SA_HVC} presents the position spectra of the \OIa and \SIIa line regions in the three years. The \OIb~line is not shown in this figure as the extended jet is faint or absent in all epochs of data for this line. The strong variability 
in the centroids beyond $\sim$ -75~kms$^{-1}$ trace the changes in the jet as the new knots are ejected with decreasing velocities. The region of the HVC1 initially shows a positive velocity gradient as material is accelerated in the jet \citep{Whelan2021}. The peaks and troughs in the region of the position spectrum that covers the jet emission are typical of knotty jets with the knots representing shocks in the outflow. For example, Figure 5 of \cite{Whelan2004} and many of the figures presented in  \cite{Hirth1997} show signatures in their position spectra consistent with knotty jets. The HVC2 is located at $\sim$ 30~au in \OIa in 2010 in agreement with the \cite{Chou2024} analysis of the 2003 HDS spectra. The region of the MVC and LVC-H in both lines also has a positive velocity gradient, linking them to the jet, although in some cases the gradient is not steep.

\subsection{Variability in the LVC}\label{Sec3.2}
The LVC-M and LVC-L kinematic components are now discussed in terms of changes in the shape of the line profile, the peak velocities measured by the kinematic fitting and the spatial properties of the emission regions as mapped with spectro-astrometry. These components are not spatially resolved in the PV diagrams of Figure \ref{DGTau_pvplots}. The top row of Figure \ref{DGTau_1dSA_compare1} shows the LVC regions of each line profile with each epoch over-plotted. The lines have been normalised to the line peaks so changes in the LVC line profile shape become clear. As can be seen from Table \ref{table_kinfit_comparison}, the peak velocities of these components do not vary greatly, unlike the jet components. The range of the LVC-M is marked in green in Figure \ref{DGTau_1dSA_compare1} and the LVC-L is marked in red, with the overlap region appearing in light brown. This brown colour highlights the strong overlap between the two components which impacts the kinematic fitting and spectro-astrometry. The line profile shape for all three lines changes substantially across the three epochs. This changes what components are detected by the kinematic fitting, the measured peak velocities and peak centroids measured by spectro-astrometry.  
\newline\par\noindent 
For \OIb\ in 2003 the line is at its widest and spans both the LVC-M and LVC-L. The LVC-M is very prominent in this epoch and therefore the kinematic fitting only detects a LVC-M. In 2010 and 2021 an LVC-L is detected by the kinematic fitting and the LVC-M is not, as what was the LVC-M has reduced to a blue-wing. The peak velocities recorded for these components are consistent with the other emission lines.
\newline\par\noindent
The \OIa LVC consists of a peak, an extended blue-wing and a less substantial red-wing. In 2003 the kinematic fitting measures the line peak at -20~kms$^{-1}$ and this is identified as the LVC-M and a hump in the red-shifted wing is measured at -9~kms$^{-1}$ and identified as the LVC-L. The shape of the line profile changes substantially in 2010 and 2021. The LVC-M is associated with a blue-wing in 2010 and a hump in the blue-wing in 2021, rather than a clear peak as in 2003. The LVC-L which appears as a hump in the red wing in 2003 is a clear line peak in 2010 and 2021. The peak velocity of the LVC-M increases from -20~kms$^{-1}$ in 2003 to -27~kms$^{-1}$ in 2010 and back to -22~kms$^{-1}$ in 2021. We argue that this velocity variation for the LVC-M in \OIa is due to some blending with the MVC jet component which is prominent in 2010 and the other changes to the line profiles described above. The peak velocity of the LVC-L does not vary by more than 2~kms$^{-1}$. 
\newline\par\noindent    
The shape of the \SIIa LVC also changes between 2003 and 2021. The red-wing seen in 2003 becomes a hump in 2010 and 2021. It could be that the LVC-L is contributing to the line profile in 2010/2021 but this contribution is not picked up by the kinematic fitting. This interpretation to the explain the changes observed in the \SIIa LVC aligns with the increase in the contribution of the LVC-L to both [O I] lines in those years. The strong and wide blue-shifted emission peak seen in 2003 and 2010 reduces to a blue-wing in 2021. The peak velocity of the LVC-M increases from -30~kms$^{-1}$ in 2003 to -40~kms$^{-1}$ in 2010 and back to -20~kms$^{-1}$ in 2021. Again, we argue that these changes to the peak velocity of the LVC-M is due to some blending with the MVC jet component which is prominent in 2010 and the other changes to the line profiles described above. 
\newline\par\noindent
The bottom row of Figure \ref{DGTau_1dSA_compare1} presents the position spectra of the LVC-M and LVC-L regions, for all three FELs, and in all three epochs, after continuum subtraction. Strikingly while as highlighted above, the shape of the LVC profiles change greatly, the position spectra in this region are very consistent across the nine measurements when we consider the velocities at which the peak centroids are measured and the velocity gradients. The biggest change in the velocity at which the maximum centroid is measured ($\sim$ -4~kms$^{-1}$) occurs in the \OIa emission region where both the LVC-M and LVC-L have comparable contributions. In contrast, no such change in the velocity of the maximum offset is observed for the \SIIa measurements, in which the contribution of the LVC-L is minimised. For \OIb\ the position spectrum is most like what is seen in \OIa and [S II]$\lambda$6731 in 2003, when the \OIb\ emission is least noisy. 
\newline\par\noindent
The velocity gradients are also remarkably consistent. A positive gradient is seen for all lines at velocities more blue-shifted than -25 kms$^{-1}$ and it marks where we begin to see blending between the LVC and jet components. A negative velocity gradient where offset increases with decreasing velocity is indicative of a disk wind and is observed in all emission lines and in all epochs for the region covering 0 to -25~kms$^{-1}$. 
\newline\par\noindent
Where we do see variability in the position spectra is in the maximum and minimum values of the spatial centroids. If the LVC-L is tracing the upper disk atmosphere it would be more compact than the extended disk wind traced by the LVC-M. As the components are blended, the measured centroids are a combination of the position of both components. Where contributions from the LVC-M and LVC-L to the overall line profile are comparable, as in [O I]$\lambda$6300, there is not much change in the maximum centroid between the years. In \SIIa the peak centroid decreases from $\sim$ 55~au in 2003 to $\sim$ 35~au in 2010/2021. This is further evidence that the change in the blue-wing of the \SIIa emission in those years is due to an increased contribution from the LVC-L. The peak centroid is less in 2010/2021 as it includes a stronger contribution from the more compact LVC-L. 
\newline\par\noindent
Also it can be noted that the spatial centroids of the \OIb\ LVC-M reach a minimum value of 2~au at -25~kms$^{-1}$ in 2003 while in \OIa they are $\sim$ 15~au in all years and vary between 20~au and 32~au in \SIIa. This is evidence that the jet components are combining with the LVC-M to increase the measured offset of the more compact LVC-M emission in a similar way to how the LVC-L decreases it at velocities close to 0~kms$^{-1}$. There is no jet emission in \OIb\ in 2003 and so this does not occur in this case. The position spectra of the \OIb\ line in 2010 and 2021 are more difficult to interpret due to the relative faintness of the emission in those years so no conclusions are made. Similarly, no conclusions are made about the position spectra of the LVC red wing as measurements are impacted here by the accuracy of the continuumn subtraction. 
\newline\par\noindent
\section{Discussion and Conclusions}\label{Sec4}
\cite{Chou2024} previously analysed the first epoch of high spectral resolution data of the FEL regions of DG~Tau included in this study using kinematic fitting and spectro-astrometry. They identified seven distinct velocity components, six of which are associated with the DG Tau blue-shifted outflow. In this article we expand on this work by looking at the changes in time of the velocity components of the blue-shifted outflow, over a period of $\approx$ 18 years using an additional two epochs of high spectral resolution data. The central goal is to compare the variability in the jet and LVC to arrive at a better understanding of the origin of the LVC. 
\newline\par\noindent
Results for the jet components reveal remarkable variability which is connected to the slowing of the DG Tau jet. The HVC1 is tracing a jet, the velocity of which decreases from 2003 to 2021. \cite{Pyo2024} correlate the changes in the velocity of the jet with changes in stellar mass accretion rate. The variability of the mass accretion rate is also linked to variability in the photometric magnitude of DG Tau. \cite{Pyo2024} outline that variable accretion causes an increase in the radius at which the high velocity jet is launched and thus a decrease in the jet velocity is seen. This is because the jet velocity is approximately proportional to the Keplerian velocity at the launch radius of the jet. 
\newline\par\noindent
\cite{Chou2024} conclude that both the HVC1 and HVC2 components of the DG~Tau FELs originate from the postshock regions in the jet, supported by the fact that these components are observed in emission lines from significantly different ionization conditions e.g. [O I] and He I. They measure the density of the HVC2 at n$_{e}$ $\sim$ 10$^{6}$~cm$^{-3}$, which is higher than the HVC1 (n$_{e}$ $\sim$ 10$^{4}$~cm$^{-3}$) and find using spectro-astrometry that it is located at $\sim$ 0\farcs2 ($\sim$ 25~au) from DG~Tau. This is all strong evidence that it is the base of the DG~Tau jet and is associated with a stationary shock component previously detected in X-rays \citep{Gudel2011}. 
The HVC2 is found in this study in the 2010 UVES spectra at a similar distance but at a lower velocity, $\sim$ -122~kms$^{-1}$ compared to -185~kms$^{-1}$ in 2003. This is further evidence that it is a shock at the base of the jet, as the jet also decreases in radial velocity by $\sim$ 70~kms$^{-1}$ over this time period. \cite{Gudel2011} also report the emission from the X-ray knot to be variable which could explain its non-detection in [O I] in some years in this work. 
\newline\par\noindent
The MVC is a another velocity component associated with the jet. Its connection to the jet is supported by the decrease in its radial velocity from 2003 to 2021 in both \OIa and [S II]$\lambda$6731 and the positive velocity gradient recorded by the spectro-astrometric analysis. We propose that the MVC is emission from the new lower velocity knots being ejected into the DG~Tau blue-shifted jet and this could be confirmed by additional epochs of data. The non-detection of this component in \OIb\ is extra evidence that it is a shock in the jet. 
\newline\par\noindent
The LVC-H is the final jet component to consider. In this study and that of \cite{Chou2024} the LVC-H is not detected in \OIb\ which points to it having a lower density and/or temperature than the components detected in \OIa and thus it is likely not coming from a region close to the star. \cite{Chou2024} discuss how gas from the wind entrained by the fast jet could explain the LVC-H and reference the detection of a similar component in the [Fe II]~1.64~$\mu$m emission from DG~Tau studied by \cite{Pyo2003}. In the observations of \cite{Pyo2003} the offset of the entrained component increases with velocity from $\sim$ -60~kms$^{-1}$ and at $\sim$ 30~au from the star. This is in agreement with the velocities of the LVC-H, the spatial centroids and the positive velocity gradient recorded in this study. 
\newline\par\noindent
While the variability in the jet components is significant due to new ejections in the jet at decreasing velocities, the changes observed in the LVC are primarily driven by varying contributions from the LVC-M and LVC-L components, which mainly affect the shape of the line profile. Figure \ref{DGTau_1dSA_compare1}, shows that the contribution from the LVC-L to the overall line shape increases from 2003 to 2021 in all lines, while the contribution from the LVC-M decreases over that period. The contribution from the LVC-L to the \SIIa line is never strong enough for it to be detected through the kinematic fitting. 
\newline\par\noindent
\cite{Chou2024} argue that the LVC-L comes from the upper disk atmosphere based on the LVC-L \OIb/\OIa line ratio. This component has the highest value for this ratio indicating that it has the highest density or temperature (see their Figure 7). Their evidence for the origin of the LVC-M in a disk wind comes from the negative velocity gradient recorded between 0 to -17~kms$^{-1}$ for their analysis of the \OIb, \OIa and \SIIa lines.  This is a strong signature of a disk wind. As outlined in \cite{Whelan2021}, this negative gradient means that wind streamlines are decreasing in velocity but increasing in height with increasing disk radius. As described in Section \ref{Sec3.2} this study measures the same negative gradient in all three years. As this measurement is made in the brown part of the position spectrum (Figure \ref{DGTau_1dSA_compare1}) the question remains as to whether the gradient is a property of the LVC-M, LVC-L or the two components combined. \cite{Chou2024} associate this signal with the LVC-M for the following reasons. Firstly, the \SIIa line has a strong negative gradient but does not have a strong contribution from the LVC-L. Secondly, the negative gradient is also strong in the 2003 \OIa line where the contribution from the LVC-L is again marginal.
Thirdly, they point out that it could not be tracing one of the higher velocity components e.g. the LVC-H, as the negative gradient is seen in \OIb\ which does not trace these components. Overall they argue that the LVC-M emission contributes a larger velocity
range within the negative velocity gradient in the position spectrum. 
\newline\par\noindent
Our analysis of the 2010 and 2021 LVC confirms all of the above points made by \cite{Chou2024}. Even when a stronger contribution from the LVC-L begins to be seen in 2010 and 2021, the gradient of the position spectrum does not change. Therefore, it is also concluded here that the negative velocity gradient is a property of the LVC-M and that the LVC-M traces a disk wind. An important result from this work is that the velocity of this wind is stable compared to the velocity of the jet. It appears that the only effect the LVC-L has on the position spectrum is to reduce the size of the centroids in line with it originating from the upper disk atmosphere.
\newline\par\noindent
The next question to be addressed is whether the LVC-M traces a MHD disk wind or a photoevaporative wind.  By comparing the line profiles and spatial scales of the LVC-M sub-component with the photoevaporative and MHD disk wind models of \cite{Weber2020}, \cite{Chou2024} conclude that neither model can fully explain the LVC-M. In \cite{Whelan2021} the minimum de-projected vertical height of the LVC sub-component tracing the disk wind, is measured to be 2.7~au. As this is well within the gravitational potential well of the star, they argue that this rules out a photoevaporative wind component in this case. In Figure \ref{DGTau_1dSA_compare1},  the LVC-M emission in 2003 \OIb\ is also traced to a minimum de-projected vertical height of 2~au above the disk mid-plane, meaning that the same argument can be used here for the LVC-M. Very interestingly the lack of change in the velocity of the LVC could be taken as evidence that it is not linked to the jet as suggested by MHD jet launching models.
\newline\par\noindent
The following are the main conclusions from this study 

\begin{enumerate}

\item Up to six blue-shifted components in the FEL line profiles alongside a red-shifted wing are identified in agreement with the study of \cite{Chou2024}. The slowing of the DG Tau jet is confirmed and a decrease in velocity of $\approx$ 100~kms$^{-1}$ is seen from 2003 to 2021.
\item Further evidence that the HVC2 is a shock at the base of the DG~Tau jet is presented. This includes a position for the HVC2 in agreement with the known location of the X-ray knot and a decrease in the velocity of the HVC2 from 2003 to 2010 of the same size as the decrease in the jet velocity.
\item The analysis of the LVC-H while limited by the spectral resolution of the data supports the idea of \cite{Chou2024} that it is material entrained by the jet. 
\item The LVC-M and LVC-L are found to be variable in brightness and their peak velocity. However, the  velocity gradient for these components does not change significantly between the years or the three lines. 
\item In agreement with \cite{Chou2024} the negative velocity gradient recorded for the region of the  LVC-M and LVC-L is associated with the LVC-M and it is concluded that this LVC sub-component is tracing a disk wind. The minimum de-projected height of LVC-M in \OIb\ of 2~au favors a MHD wind. \item It is particularly noteworthy that the velocity of the LVC-M does not change as the velocity of jet decreases.  
\end{enumerate}

Overall this study shows the potential of combining kinematic fitting and spectro-astrometry to track variability and examine the origin of LVC sub-components. However, there are some key caveats. For a complex system like DG~Tau there are limitations to the kinematic fitting  due to blending between velocity components. The low time cadence of our data also limits our ability to assess if there is a time lag between the changes observed in the different velocity components in the outflow. However, these results do provide some very interesting new constraints for models of outflows emitted from YSOs. The results of this work especially highlight the need for future observational studies to investigate these outflows at a much higher cadence sampling in time (e.g. observations every few months instead of years). In particular this would be used to explore the possibility of a time lag between changes in the high velocity jet and changes in the low velocity wind in an MHD disk wind scenario.

\begin{acknowledgements}
      N. Otten acknowledges funding from the Maynooth University Graduate Teaching Scholarship and Taighde Éireann (Research Ireland) under the RI-ESO Studentship Agreement for this work. 
\end{acknowledgements}

\bibliographystyle{aa}
\bibliography{ref}

\end{document}